\documentclass[prl,aps,amsmath,twocolumn,amssymb,showpacs,floats]{revtex4}
\usepackage{epsfig,bm}
\begin{document}

\title{Modal Extraction in Spatially Extended Systems}
\author{Kapilanjan Krishan}
\affiliation{Department of Physics and Astronomy, University of California - Irvine, Irvine, California 92697}
\author{Andreas Handel}
\affiliation{Department of Biology, Emory University, Atlanta, Georgia 30322}
\author{Roman O. Grigoriev}
\author{Michael F. Schatz}
\affiliation{Center for Nonlinear Science and School of Physics, Georgia 
Institute of Technology, Atlanta, Georgia 30332}
\date{\today}

\begin{abstract}

We describe a practical procedure for extracting the spatial structure and the
growth rates of slow eigenmodes of a spatially extended system, using a unique
experimental capability both to impose and to perturb desired initial 
states. The procedure is used to construct experimentally the spectrum of
linear modes near  the secondary instability boundary in 
Rayleigh-B\'{e}nard convection.  This technique suggests an approach to 
experimental characterization of more complex dynamical states such as
periodic orbits or spatiotemporal chaos.  

\end{abstract}

\maketitle

Numerous nonlinear nonequilibrium systems in nature and in technology 
exhibit complex behavior in both space and time ; understanding 
and characterizing such 
behavior (spatiotemporal chaos) is a key unsolved problem in nonlinear science \cite{cross}.  
Many such systems are modelled by partial differential equations; hence, in 
principle, their dynamics takes place in an infinite dimensional phase space.  
However, dissipation often acts to confine these systems' asymptotic 
behavior to finite-dimensional subspaces known as invariant manifolds 
\cite{manneville}.  Knowledge of the invariant manifolds provides a wealth
of dynamical information; thus, devising methodologies to determine 
invariant manifolds from experimental data would significantly advance understanding of spatiotemporal chaos.

\begin{figure}
\includegraphics[width=8cm]{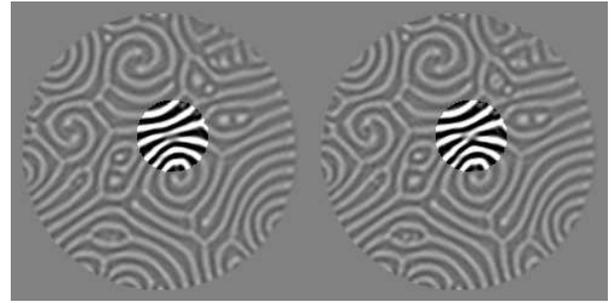}
\caption{\label{eps:sdc}
Shadowgraph visualization reveals spontaneous defect nucleation in the spiral defect chaos state of  Rayleigh-Benard convection. Two convection rolls are compressed together (higher contrast region in left image).  (b.) A short time later (right image), one of the rolls pinches off and two dislocations form.} 
\end{figure}

In this Letter, we describe experiments in Rayleigh-B\'enard convection 
where several slow eigenmodes and their growth rates associated with
instability of roll states are extracted quantitatively.  Rayleigh-B\'enard
convection (RBC) serves well as a model spatially extended system; in 
particular, the spiral defect chaos (SDC) state in RBC is considered 
an outstanding example of spatiotemporal chaos.  In SDC the spatial 
structure is primarily composed of curved but 
locally parallel rolls, punctuated by defects 
(Fig. \ref{eps:sdc}) \cite{morris,egolf1}.  The recurrent formation and 
drift of defects in SDC is believed to play a key role in driving 
spatiotemporal chaos; moreover, many aspects of defect nucleation in SDC are 
related to defect formation observed at the onset of instability in patterns 
of straight, parallel rolls in RBC \cite{busse}.   We obtain experimentally
a low-dimensional description of the modes responsible for the 
nucleation of one important class of defects (dislocations), by first 
imposing reproducibly a linearly stable, straight roll state 
(stable fixed point) near instability onset.  
This state is subsequently subjected to a set of distinct, 
well-controlled perturbations, each of which initiates a relaxational 
trajectory  from the disturbed state to the (same) fixed point.   
An ensemble of such trajectories is used to 
construct a suitable basis for describing the embedding space by  means of a
modified Karhunen-Loeve decomposition.  The dynamical evolution of small disturbances
is then characterized by computing both finite-time Lyapunov exponents and 
the spatial structure  of the associated eigenmodes (a similar approach was carried 
out numerically by {\em Egolf et al.} \cite{egolf2}).   This capability is an important
step toward developing a systematic way of characterizing and, perhaps, controlling, 
spatiotemporally chaotic states like SDC where  localized ``pivotal'' events 
like defect formation play a central role in driving complex behavior.



The convection experiments are performed with gaseous CO$_2$ at a
pressure of 3.2 MPa. A 0.697$\pm$0.06 mm-thick gas layer is contained 
in a 27 mm square cell, which is 
confined laterally by filter paper.  The layer 
is bounded on top by a sapphire window and on the bottom by a
sheet of 1 mm-thick glass neutral density filter(NDF).  The neutral density
filter is bonded to a heated metal plate with heat sink compound.
The temperature of the sapphire window held constant at 21.3 $^{\circ}$C by
water cooling.  The temperature difference between the top and
bottom plates $\Delta T$ is held fixed at 5.50 $\pm$ 0.01 $^{\circ}$C 
by computer control of a thin film
heater attached to the bottom metal plate.  
These conditions correspond to 
a dimensionless bifurcation
parameter $\epsilon$=$(\Delta T - \Delta T_c)/\Delta T_c=0.41$, where $\Delta
T_c$ is the temperature difference at the onset of convection. The vertical thermal diffusion time, computed to be 2.1 s at onset, represents the characteristic timescale for the system.

We use laser heating to alter the convective patterns that occur
spontaneously. A focused beam from an Ar-ion laser is directed through the
sapphire window at a spot on the NDF. Absorption of the laser light by
the NDF increases the local temperature of the bottom boundary and hence that
of the gas, thereby inducing locally a convective upflow.  The convection
pattern may be modifed, either locally or globally, by rastering the hot
spot created by the laser beam. The beam is steered using two galvanometric
mirrors rotating about axes orthogonal to each other under computer control.
The intensity of the beam is modulated using an acousto-optic modulator. This
technique of optical actuation is used to impose convection
patterns with desired properties, to perturb these convection
patterns and to change the boundary conditions. 
Similar approaches for manipulating
convective flows were explored earlier using a high intensity lamp and masks
\cite{whitehead} in RBC and a rastered infrared laser in B\'{e}nard-Marangoni
convection \cite{denis}.

\begin{figure}
\begin{center}
{\includegraphics[width=4cm]{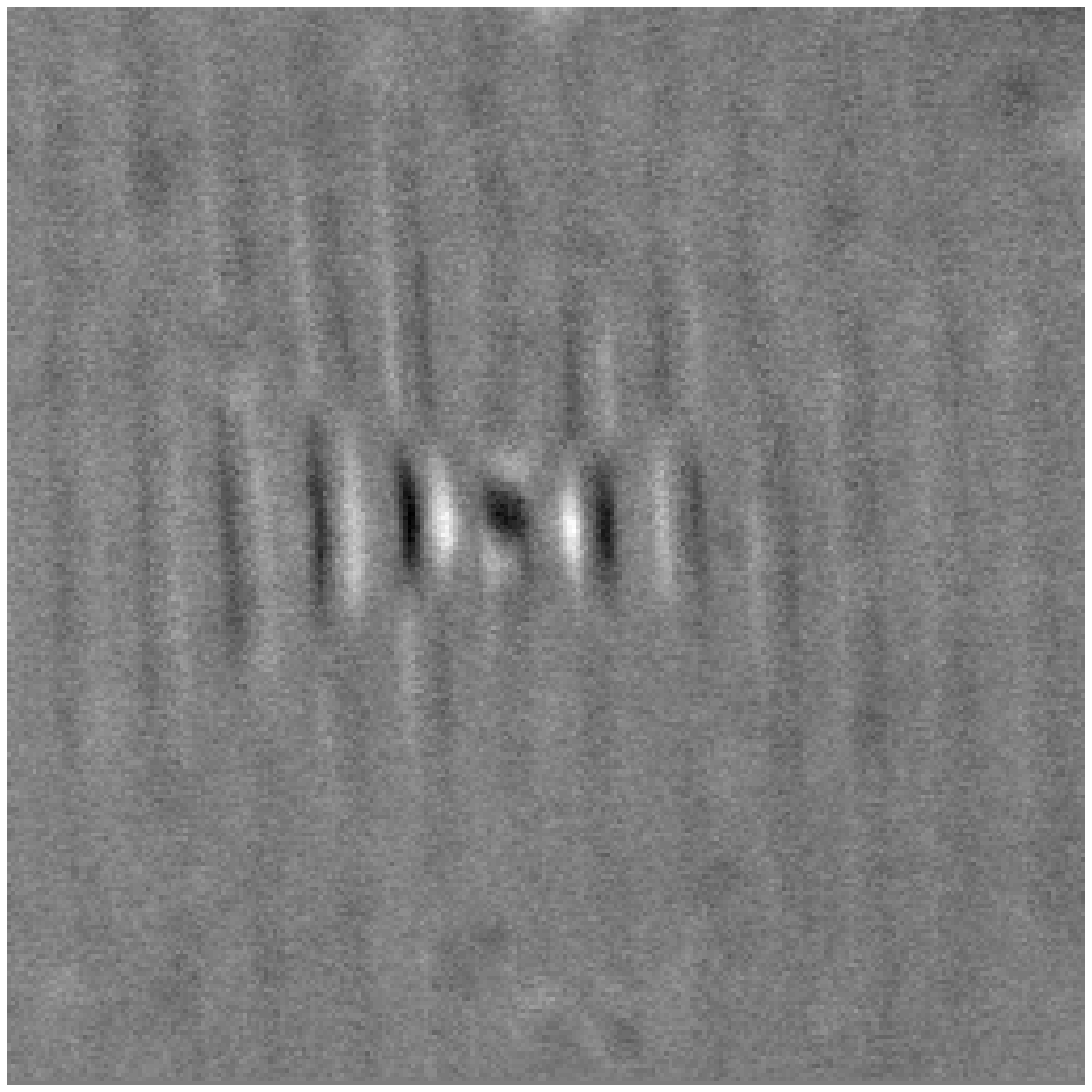}} 
\hspace{4pt}%
{\includegraphics[width=4cm]{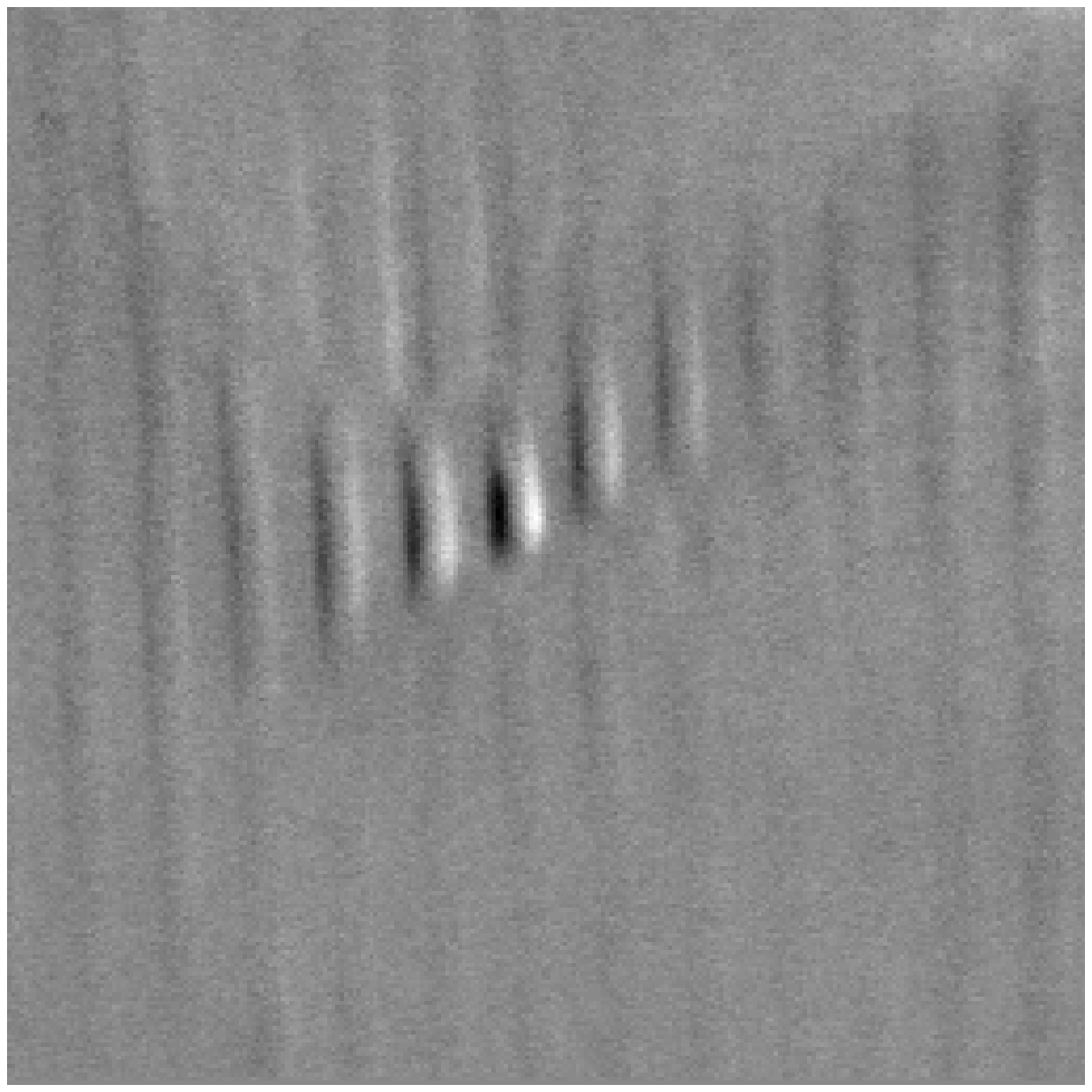}} 
\end{center}%
\vspace{-3mm}
\caption{\label{eps:rd}
Experimental images illustrate the flow response to two different 
perturbations applied, in turn, to the same state of straight 
convection rolls.  Each image represents
the difference between the perturbed and unperturbed convection states and
therefore, each image highlights the effect of a given perturbation
on the flow.  In the two cases shown, the localized perturbation is 
applied directly on a
region of  either downflow (left image)  or upflow (right image).  In all cases,
the disturbance created by the perturbation decays away and the flow
returns to the original unperturbed state.}
\end{figure}

The experiments begin by using laser heating to impose a well-specified 
basic state of stable straight rolls.  
The basic state is typically arranged to be near the onset of instability 
by imposing a sufficiently large 
pattern wavenumber such that at fixed $\epsilon$ the system's 
parameters are near the
skew-varicose stability boundary \cite{busse}.   In this regime, the modes
responsible for the instability are weakly damped and, therefore, can be 
easily excited.

The linear stability of the basic state is probed by applying brief pulses of
spatially localized laser heating. 
For stable patterns, all small disturbances 
eventually relax. 
To excite all modes governing the disturbance evolution, we apply a set
of localized perturbations consistent with symmetries of the (ideal) straight roll
pattern -- continuous translation symmetry in the direction along the rolls and
discrete translation symmetry in the perpendicular direction plus the
reflection symmetry in both directions. Therefore, localized perturbations applied
across half a wavelength of the pattern form a "basis" for all such perturbations --
any other localized perturbation at a different spatial location is related by symmetry. Localized perturbations
are produced in the experiment by aiming the laser beam to create 
a  ``hot spot'' 
whose location is stepped from the center of a (cold) downflow region to the center of an adjacent (hot) upflow region in
different experimental runs.  The perturbations typically last approximately 5 s and have a lateral extent of approximately 0.1 mm, which is less than
10 \% of the pattern wavelength.

The evolution of the perturbed convective flow is monitored by 
shadowgraph visualization. 
A digital camera with a low-pass filter 
(to filter out the reflections from the Ar-ion
laser) is used to capture a sequence of $256\times 256$ pixel images recorded 
with 12 bits of intensity resolution at a rate of 41 images per second.
A background image of the unperturbed flow is subtracted from
each data image; such sequences of difference images comprise the time 
series representing the evolution of the perturbation (Fig ~\ref{eps:rd}).

The total power for each (difference) image in a time series is obtained 
from 2-D spatial Fourier transforms.  The resulting time series
of total power shows a strong transient excursion (corresponding to 
the initial response of the convective flow to a localized
perturbation by laser heating) followed by exponential decay as the
system relaxes back to the stable state of straight convection rolls.
We restrict further analysis to the region of exponential decay, which 
typically represents about $3.5$ seconds of data for each applied perturbation.

The dimensionality of the raw data is too high to permit direct analysis, so 
each difference image is first windowed 
(to avoid aliasing effects) and Fourier filtered by
discarding the Fourier modes outside a $31\times 31$ window centered at the
zero frequency. The discarded high-frequency modes are strongly damped and
contain less than 1\% of the total power. The basis of $31^2$ Fourier modes
still contains redundant information, so we further reduce the
dimensionality of the embedding space by projecting the disturbance 
trajectories onto the
``optimal'' basis constructed using a variation of 
the Karhunen-Loeve (KL)
decomposition \cite{holmes,sirovich}.  The correlation matrix $C$
is computed using the Fourier filtered time series ${\bf x}^s(t)$,
\begin{equation}
C=\sum_{s,t}({\bf x}^s(t)-\langle{\bf x}^s(t)\rangle_t)
({\bf x}^s(t)-\langle{\bf x}^s(t)\rangle_t)^\dagger,
\end{equation}
where the index $s$ labels different initial conditions and the origin of 
time $t=0$ corresponds to the time when the perturbation applied by the
laser is within the linear neighbourhood of the statioary state. The angle brackets with the subscript $t$ indicate a time average. The eigenvectors of $C$ are the KL basis vectors.  It is worth noting that the average 
performed to compute $C$ represents an ensemble average over different 
initial conditions (obtained by applying different perturbations); this is
distinctly different from the standard implementation of KL decomposition
where statistical time averages are typically employed.

\begin{figure}
\begin{center}
{\includegraphics[width=4cm]{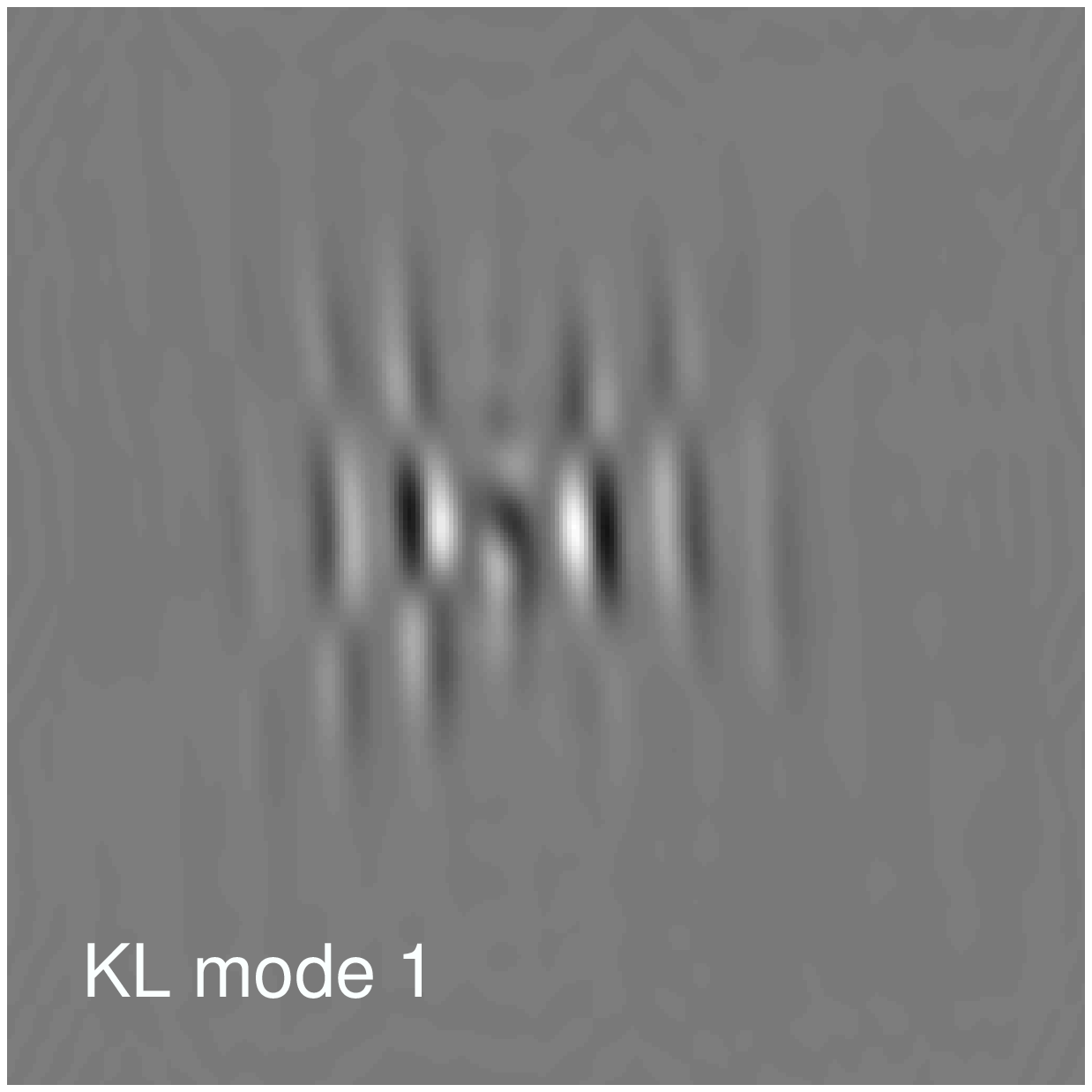}} 
\hspace{4pt}%
{\includegraphics[width=4cm]{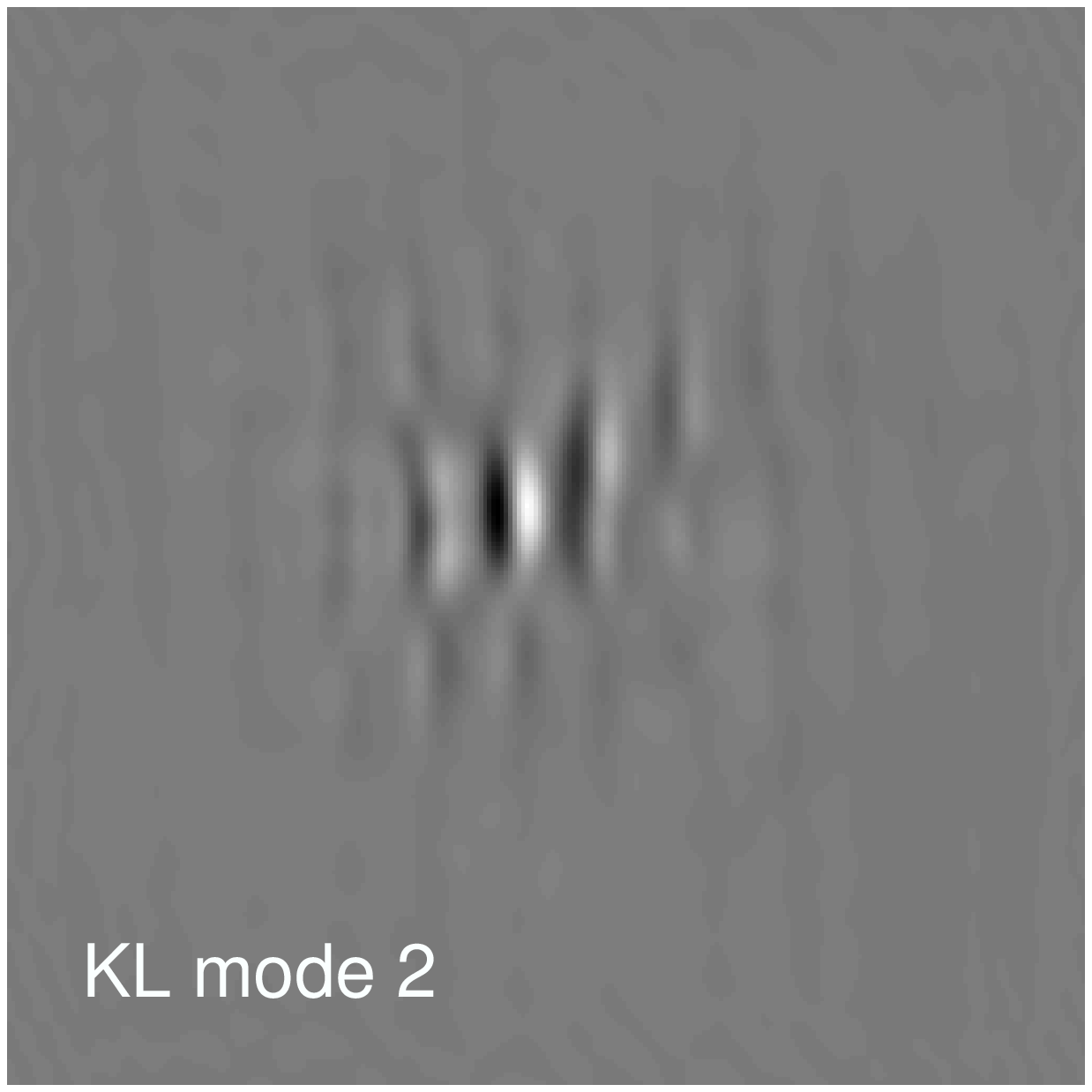}} \\
\vspace{8pt}%
{\includegraphics[width=4cm]{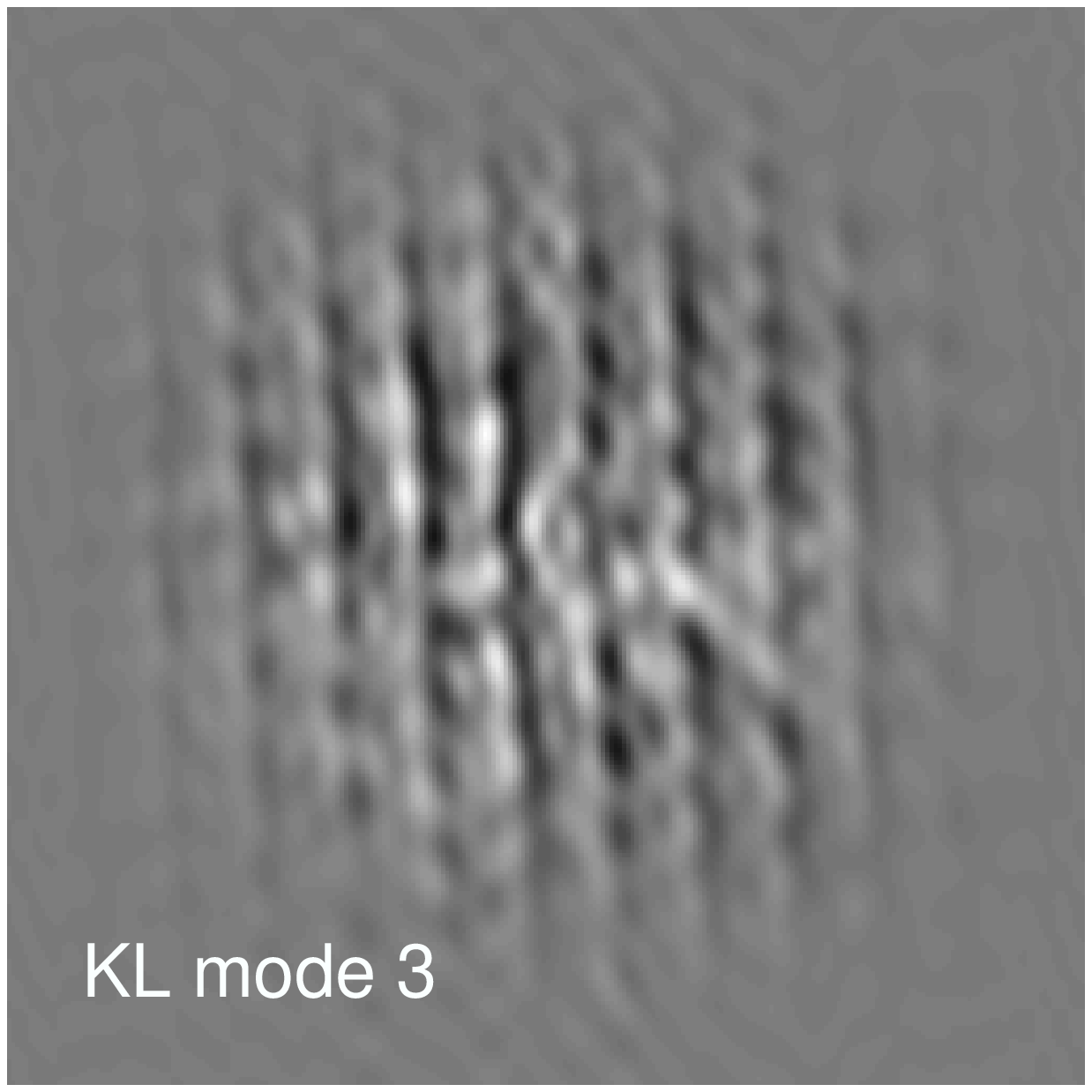}}
\hspace{4pt}
{\includegraphics[width=4cm]{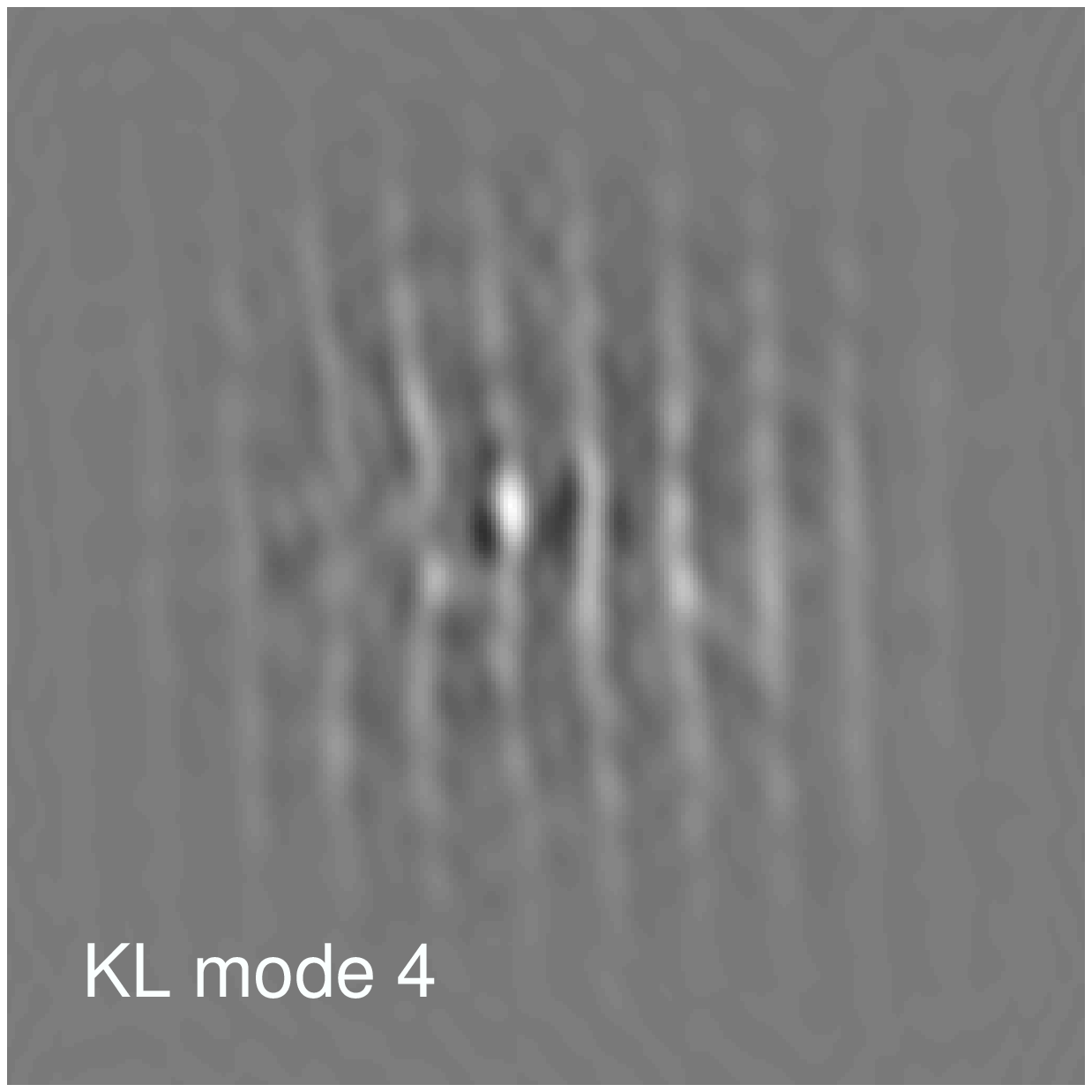}}\\
\caption{\label{eps:kl_new_stable}
The first four Karhunen-Loeve eigenvectors 
are shown for a perturbed roll state near
the skew-varicose boundary of Rayleigh-B\'enard convection.  The eigenvectors
are ordered by their eigenvalues (largest to smallest), which
are propotional to the amount of power contained in the corresponding eigenvector.}
\end{center}%
\end{figure}

The
spatial structures of the first four KL
vectors are shown in Fig. \ref{eps:kl_new_stable}. We find that the first 24
basis vectors capture over 90\% of the total power, so an embedding space
spanned by these vectors represents well the relaxational dynamics about the
straight roll pattern. In our convection experiments, the KL eigenvectors
show two distinct length scales. The first two dominant vectors are spatially
localized, while the remaining vectors are spatially extended. 
This is consistent with earlier work as suggested in \cite{egolf1}.

More quantitative information can be obtained by finding the eigenmodes of the
system, excited by the perturbation, and their growth rates. These can be
extracted from a nonlinear least squares fit with the cost function
\begin{equation}
E_n=\sum_{i,s,t}\left[{\bf x}^s_i(t)-\left({\bf x}^s_i(\infty)
+\sum_{k=1}^n A^s_k{\bf m}^k_i e^{\lambda_k t}\right)\right]^2,
\end{equation}
where ${\bf x}^s_i(t)$ is a projection of the perturbation at time $t$ in the
time series $s$ onto the $i$th KL basis vector. In the fit ${\bf m}^k$
and $\lambda_k$ are the $k$th eigenmode and its growth rate and $A^s_k$ is the
initial amplitude of the $k$th eigenmode excited in the experimental time
series $s$. The fixed points ${\bf x}^s(\infty)$ are chosen to be different for
the differing time series in the ensemble to account for a slow drift in the
parameters and we assume that  only $n$ eigenmodes are excited.

\begin{figure}[t]
\includegraphics[width=7.5cm]{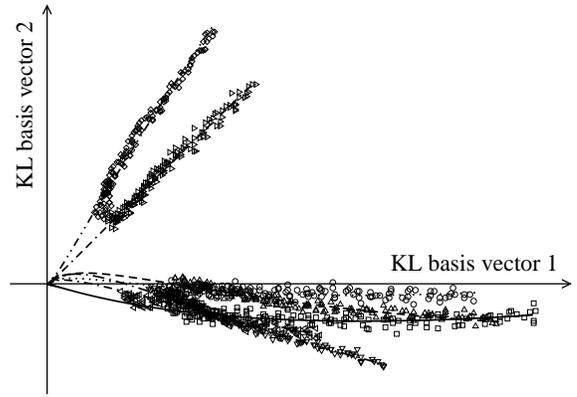}
\caption{\label{eps:ps_new_stable}
A two-dimensional projection of the experimental time series (symbols) and the
least squares fits (continuous curves). The time series have been shifted such
that the fixed point is at the origin. }
\end{figure}

\begin{figure}[t]
\includegraphics[width=8cm]{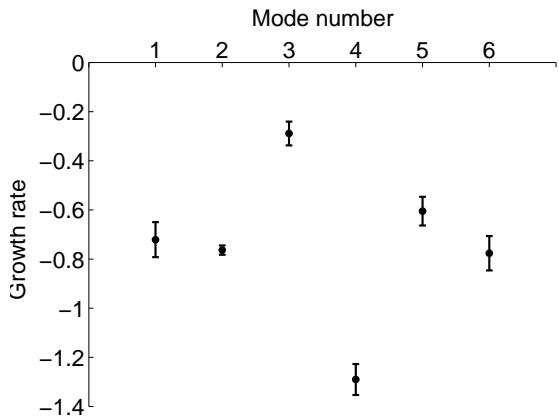}
\caption{\label{eps:gr_new_stable}
The growth rates of the six dominant eigenmodes and the error bars extracted
from the least squares fit. The growth rates have been non-dimensionalized by
the vertical thermal diffusion time.}
\end{figure}

The results for an ensemble of time series corresponding to seven point
perturbations applied across a wavelength
%
%
of the pattern with $n=6$ are shown
in Figs. \ref{eps:ps_new_stable}-\ref{eps:gr_new_stable}. (With seven
different initial conditions we cannot hope to distinguish more than seven
different modes). In particular, Fig. \ref{eps:ps_new_stable} shows the
projection of the experimental time series and the least squares fit on the
plane spanned by the first two KL basis vectors. Such extraction of the linear manifold in experiments 
on spatially extended systems without the knowledge of the dynamical equations 
of the system aids in the application of techniques that are well developed 
for low dimensional systems. The manifolds of fixed points and periodic orbits 
are of particular interest in chaotic systems.

\begin{figure}[t]
\begin{center}
{\includegraphics[width=4cm]{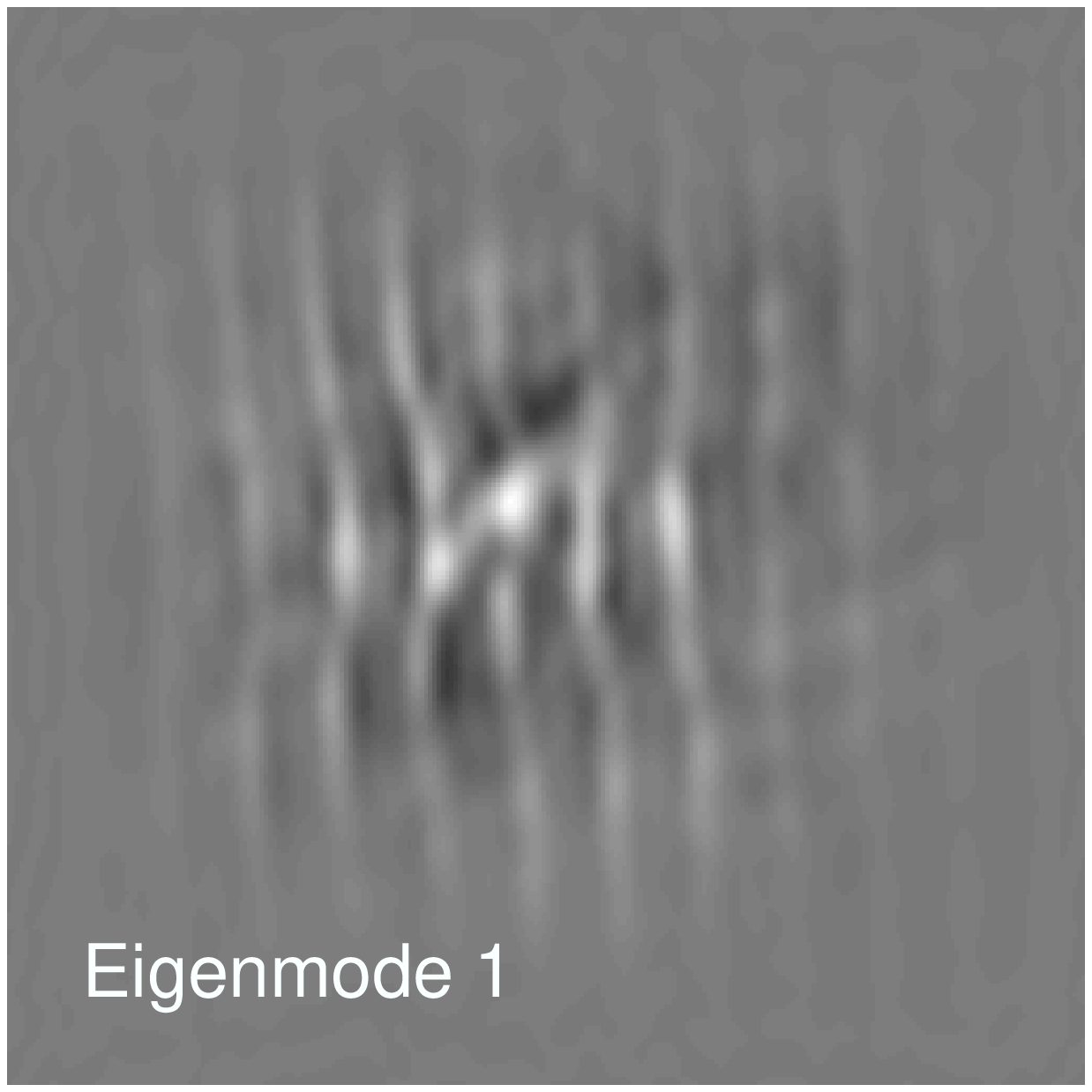}} 
\hspace{4pt}%
{\includegraphics[width=4cm]{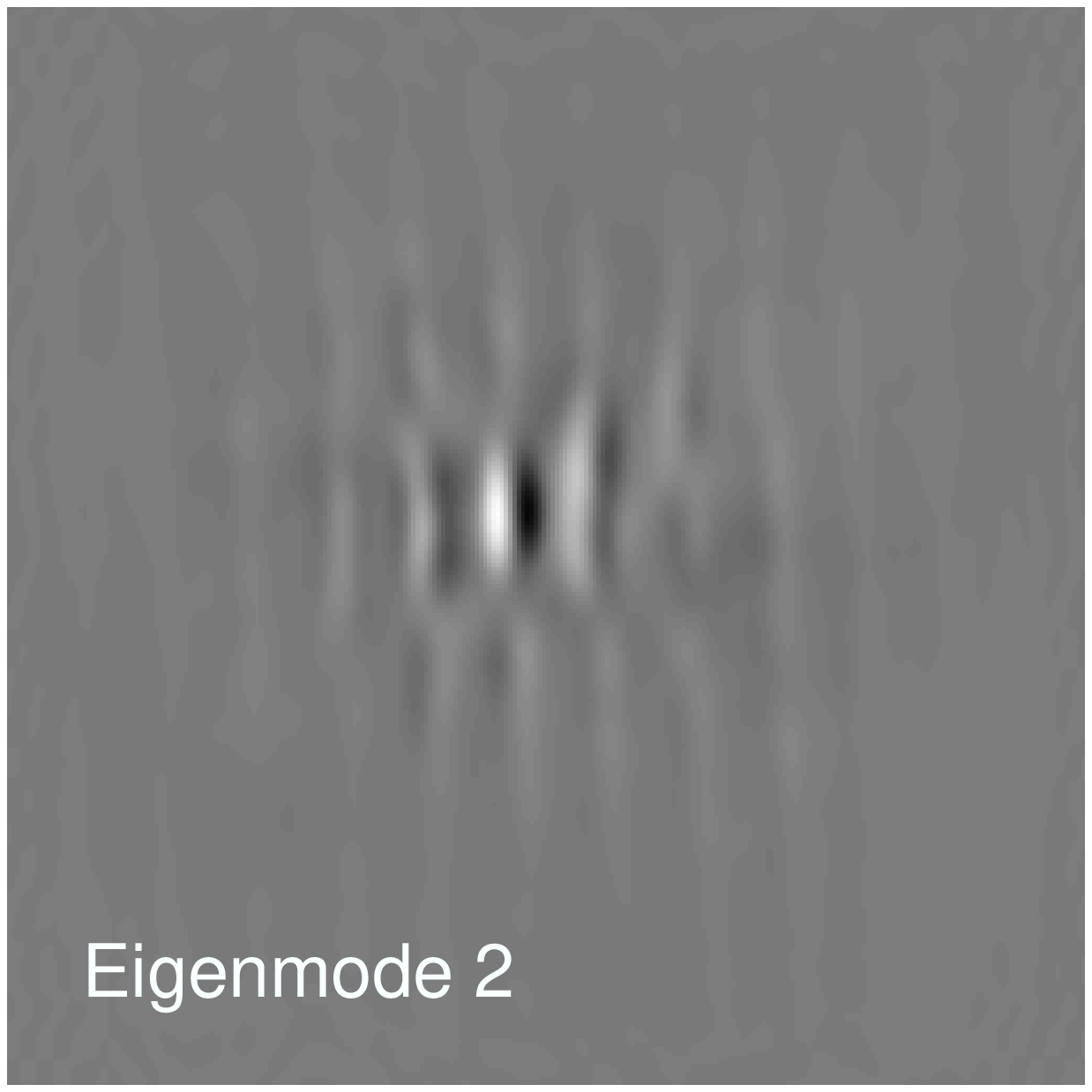}} \\
\vspace{8pt}%
{\includegraphics[width=4cm]{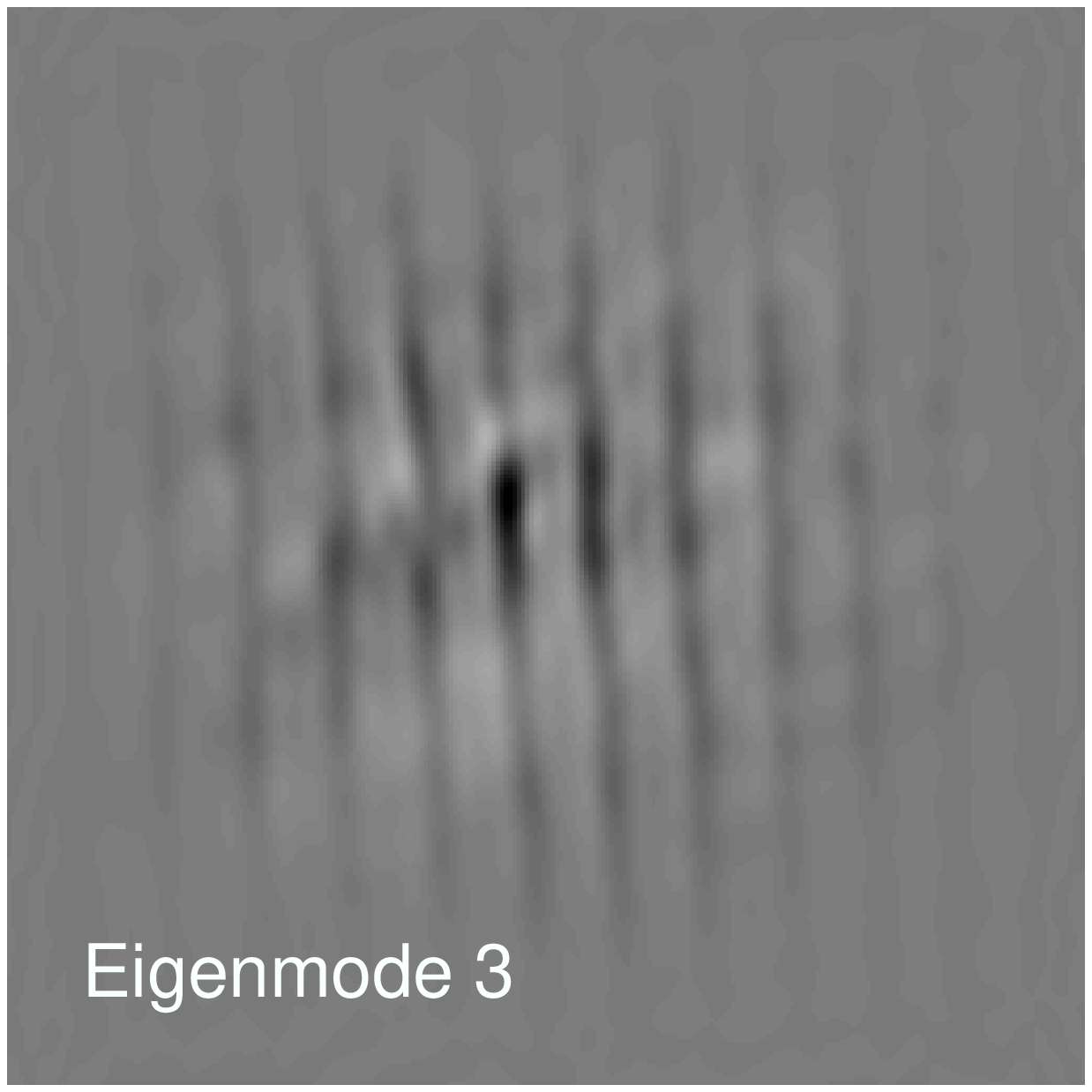}}
\hspace{4pt}
{\includegraphics[width=4cm]{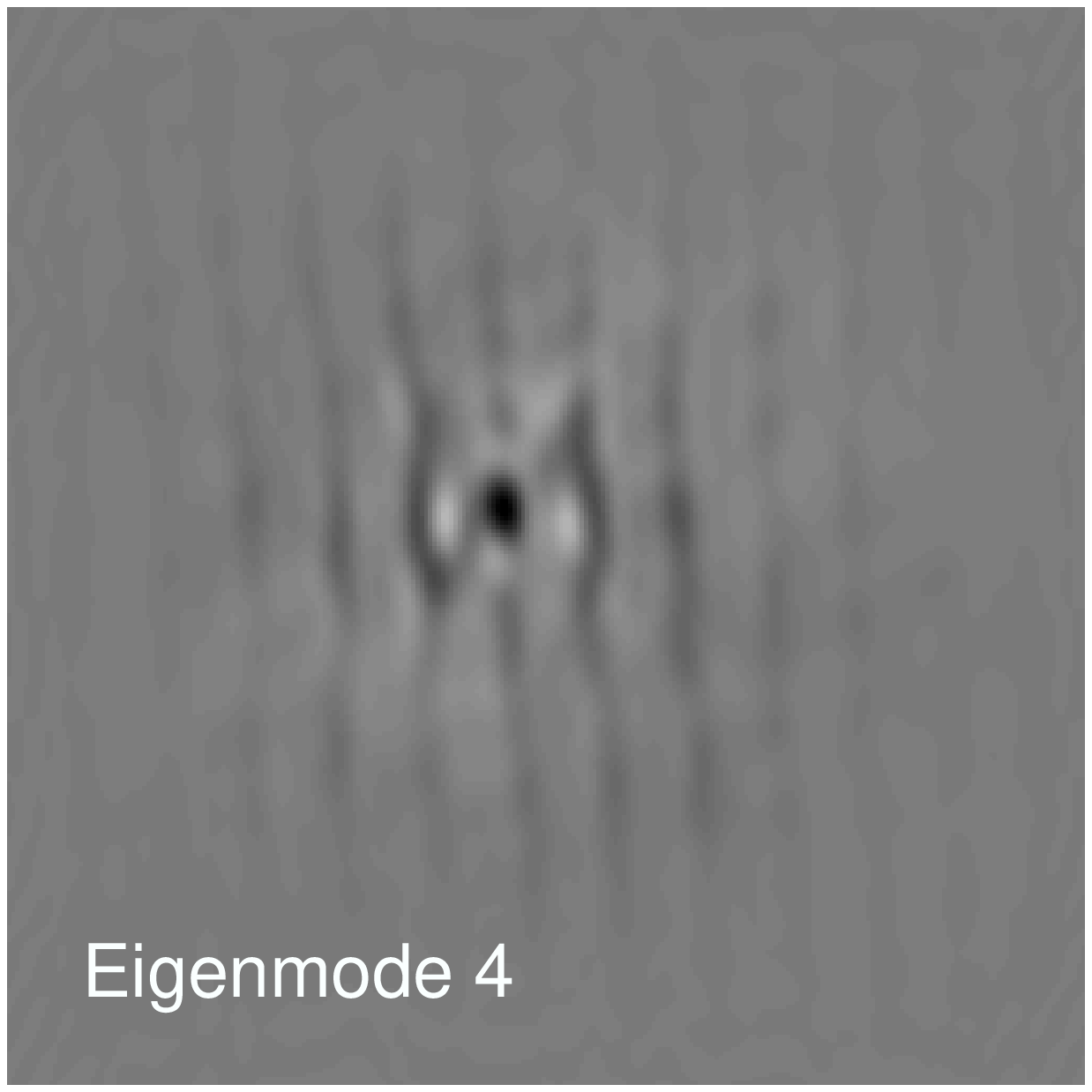}}\\
\caption{\label{eps:modes_new_stable}
Four dominant eigenmodes extracted from the least squares fit.} 
\end{center}%
\end{figure}

The extracted growth rates $\lambda_k$ are shown in Fig.
\ref{eps:gr_new_stable}. Not surprisingly, since the pattern is stable 
the growth
rates are negative. The leading eigenmode (see Fig. \ref{eps:modes_new_stable})
is spatially extended and shows a diagonal structure characteristic of the
skew-varicose instability in an unbounded system. This is also expected as the
pattern is near the skew-varicose instability boundary. The second eigenmode is
spatially localized and has no analog in spatially unbounded systems. The
subsequent modes are again spatially delocalized and likely correspond to the
Goldstone modes of the unbounded system (e.g., overall translation of the
pattern) which are made weakly stable due to confinement by the lateral 
boundaries of the convection cell.

If the system is brought across the stability boundary, one of the modes is
expected to become unstable (without significant change in its spatial
structure), thereby determining further (nonlinear) evolution of the system
towards a state with a pair of dislocation defects. We would also expect the
spatially localized eigenmodes (like the second one in Fig.
\ref{eps:modes_new_stable}) to preserve their structure if the base state is
smoothly distorted (as it would be, e.g., in the SDC state shown in Fig.
\ref{eps:sdc}), indicating the same type of a spatially localized instability.
Our further experimental studies will aim to confirm these expectations.

Defects represent a type of ``coherent structure''  in spiral defect 
chaos.  Previous efforts have used coherent structures to 
characterize spatiotemporally chaotic extended systems in both 
models \cite{sirovich}  and experiments \cite{wolf}; the use of 
coherent structures to parametrize the invariant manifold was
pioneered by Holmes {\em et al.} \cite{holmes} in the context of turbulence.
In practice coherent structures are usually extracted using 
the Karhunen-Lo\'eve (or
proper orthogonal) decomposition of time series of system states, which
picks out the {\em statistically} important patterns.  
This prior work has met
with only limited success -- indeed, it is unclear whether statistically
important patterns are {\em dynamically} important.   
An alternative approach has
been proposed by Christiansen {\em et al.} \cite{christiansen}, who suggested
instead to use the recurrent patterns corresponding to the low-period unstable
periodic orbits (UPO) of the system, which are dynamically more important.
Our work sets the stage for attempting the more ambitious task of 
extraction of UPOs and their stability properties from experimental data.

Summing up, we have developed an experimental technique which allows extraction
of quantitative information describing the dynamics and stability of a pattern
forming system near a fixed point. This technique should be applicable to a
broad class of patterns, including unstable fixed points, periodic orbits and
segments of chaotic trajectories. Moreover, we expect that a similar approach
could be applied to other pattern forming systems, convective or otherwise, as
long as a method of spatially distibuted actuation of their state can be
devised.

\end{document}